\documentclass[aps,twocolumn,pre]{revtex4-2}

\usepackage{xcolor,hyperref}
\hypersetup{
   colorlinks,
   linkcolor={blue!50!black},
   citecolor={blue!50!black},
   urlcolor={blue!80!black}
}

\usepackage{epsfig, amsmath, mathrsfs, physics}
\usepackage{bm}
\usepackage{soul}
\usepackage{hyperref}
\usepackage{footmisc}

\newcommand{\dbar}{{\,\mathchar'26\mkern-12mu d}}

\newcommand{\Tx}{T_{\mbox{\tiny$\mathsf{X}$}}}

\DeclareMathAlphabet{\mathitbf}{OML}{cmm}{b}{it}

\newcommand{\xv}{\mathitbf x}

\newcommand{\kv}{\mathitbf k}
\newcommand{\mv}{\mathitbf m}

\begin{document}

\title{A unified quantifier of mechanical disorder in solids}
\author{Geert Kapteijns$^{1}$}
\email{g.h.kapteijns@uva.nl}
\author{Eran Bouchbinder$^{2}$}
\email{eran.bouchbinder@weizmann.ac.il}
\author{Edan Lerner$^{1}$}
\email{e.lerner@uva.nl}
\affiliation{$^{1}$Institute of Theoretical Physics, University of Amsterdam, Science Park 904, 1098 XH Amsterdam, the Netherlands\\
$^{2}$Chemical and Biological Physics Department, Weizmann Institute of Science, Rehovot 7610001, Israel}

\begin{abstract}
Mechanical disorder in solids, which is generated by a broad range of physical processes and controls various material properties, appears in a wide variety of forms. Defining unified and measurable dimensionless quantifiers, allowing quantitative comparison of mechanical disorder across widely different physical systems, is therefore an important goal. Two such coarse-grained dimensionless quantifiers (among others) appear in the literature, one is related to the spectral broadening of discrete phononic bands in finite-size systems (accessible through computer simulations) and the other is related the spatial fluctuations of the shear modulus in macroscopically large systems. The latter has been recently shown to determine the amplitude of wave attenuation rates in the low-frequency limit (accessible through laboratory experiments). Here, using two alternative and complementary theoretical approaches linked to the vibrational spectra of solids, we derive a basic scaling relation between the two dimensionless quantifiers. This scaling relation, which is supported by simulational data, shows that the two apparently distinct quantifiers are in fact intrinsically related, giving rise to a unified quantifier of mechanical disorder in solids. We further discuss the obtained results in the context of the unjamming transition taking place in soft sphere packings at low confining pressures, in addition to their implications for our understanding of the low-frequency vibrational spectra of disordered solids in general, and in particular those of glassy systems.
\end{abstract}

\maketitle

\section{Introduction}
\label{sec:Intro}

Mechanical disorder in solids appears in a multitude of forms, e.g., manifested in the material's composition, in the spatial arrangement of its constituents and in the interactions between them. It can be generated by a broad range of physical processes, taking place either during solid formation (e.g., solidification or glass transition) or/and after it (e.g., through various heat treatments, irreversible mechanical deformation and irradiation). Mechanical disorder has deep impact on the properties of solids, such as stress relaxation, sound attenuation, thermal conductivity, plastic deformability and failure resistance. Consequently, quantifying mechanical disorder is important; in particular, it is highly desirable to define measurable dimensionless and universally applicable quantifiers of mechanical disorder, which allow to quantitatively compare the degree of disorder of widely different physical systems. 

\begin{figure*}[!ht]
\centering
\includegraphics[width = 1.0\textwidth]{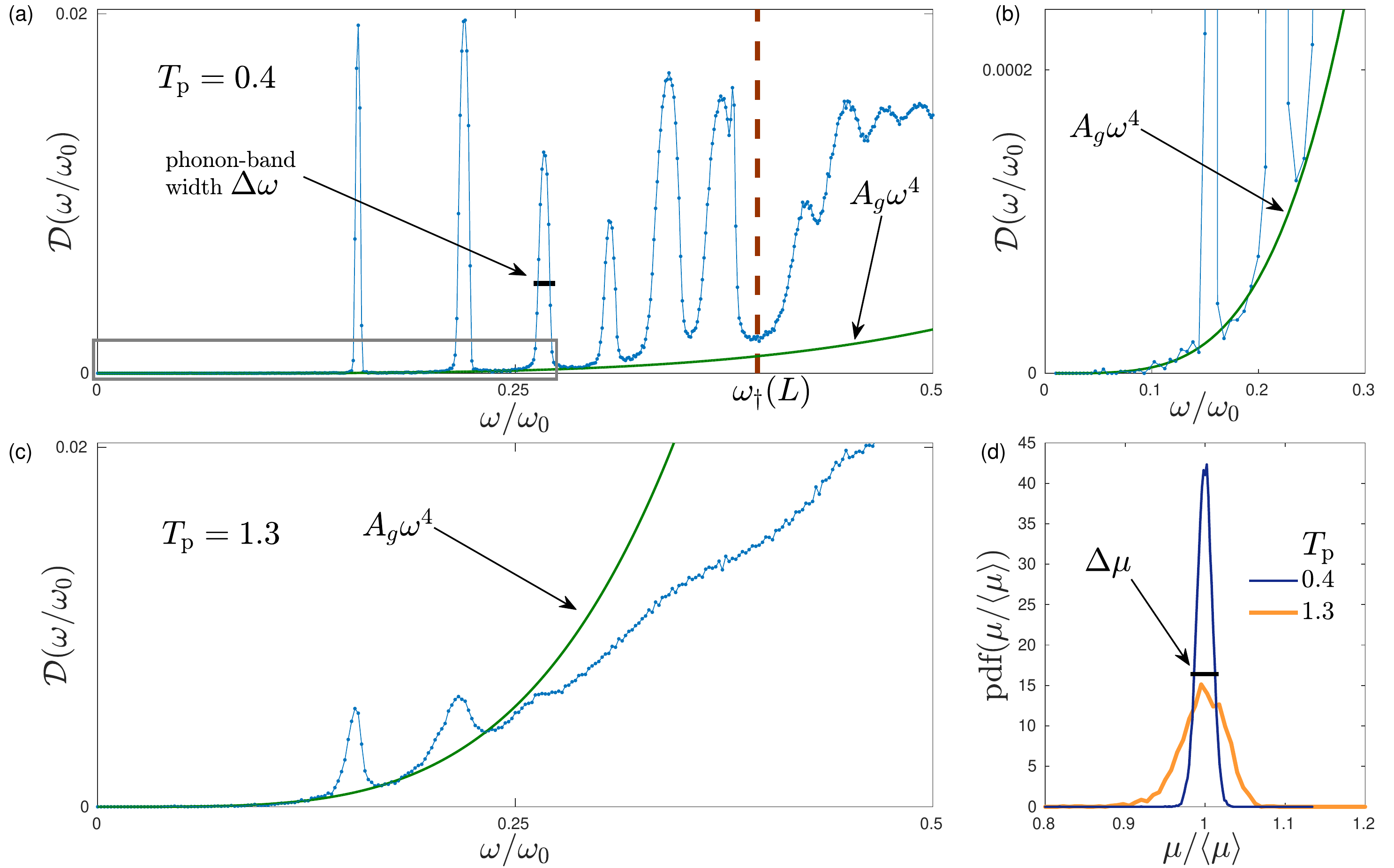}
\caption{\footnotesize (a)-(c) Low-frequency density of states ${\cal D}(\omega)$ measured in computer glasses of $N\!=\!\mbox{64,000}$  particles in three dimensions, plotted against the rescaled frequency $\omega/\omega_0$, where $\omega_0\!\equiv\!c_s/a_0$ with $c_s$ the ($T_{\rm p}$-dependent) speed of shear waves, and $a_0\!\equiv\!V/N$ is an interparticle distance. (a)~Data for glasses of parent temperature $T_{\rm p}\!=\!0.4$ (in simulational units); the frequency scale $\omega_\dagger(L)$ indicated by the dashed vertical line separates the frequency axis into the range in which discrete phonon bands are distinguishable, and the range in which their width surpasses the gaps between them, rendering them indistinguishable (`phonon sea'). The grey rectangle indicates the zoomed-in window presented in panel~(b), where the universal $\sim\!\omega^4$ VDoS of nonphononic modes is observed between and below the discrete phonon bands. (c)~Same as (a) but for $T_{\rm p}\!=\!1.3$ glasses of larger mechanical disorder. The dimensionless prefactor $A_{\rm g}\omega_0^5$ is roughly 35 times larger compared to the $T_{\rm p}\!=\!0.4$ glasses. (d) Sample-to-sample probability distribution functions (pdfs) of the shear modulus $\mu$, measured for $T_{\rm p}\!=\!1.3$ and $T_{\rm p}\!=\!0.4$ glasses of $N\!=\!$ 16,000 particles.}
\label{fig:illustration_fig}
\end{figure*}

Several proposals of quantifiers of \emph{mechanical} disorder exist in the literature~\cite{Schirmacher_2006,Schirmacher_prl_2007,schirmacher2011comments,schirmacher_ruocco_arXiv_2020,Schirmacher2021_disorder_classification,widmer2008irreversible,Hua_vibrality,alessio_inversion_symmetry,cge_paper,phonon_widths,pinching_pnas,david_indicators_PRM2020,karina_sticky1}. Of particular relevance in the present context are measurable disorder quantifiers that can be applied in both computer simulations --- that are playing increasingly important roles in materials research due to the dramatic rise in computing power --- and in laboratory experiments. As such, the disorder quantifiers we consider are coarse-grained to some extent, probing the effect of disorder on some measurable physical properties. One such quantifier can be constructed using the spatial fluctuations of the shear modulus $\mu$, denoted by $\Delta\mu$. The ratio $\Delta\mu/\langle\mu\rangle$ (where $\langle\mu\rangle$ is the average shear modulus, cf.~Fig.~\ref{fig:illustration_fig}) can be used to define a dimensionless disorder quantifier. Such a definition can make physical sense only if it is scale-independent, i.e.~if the fluctuations $\Delta\mu$ are probed on a lengthscale $\ell$ such that $(\Delta\mu/\langle\mu\rangle)(\ell/a_0)^{\dbar/2}$ is independent of $\ell$. Here $\dbar$ is the spatial dimension, $a_0$ is an atomistic length and $\ell$ is larger than any possible correlation length associated with the fluctuations of the shear modulus $\mu$. This disorder quantifier plays a central role in a class of theoretical approaches collectively termed Fluctuating Elasticity Theory~\cite{Schirmacher_2006,Schirmacher_prl_2007,schirmacher2011comments,schirmacher_ruocco_arXiv_2020,Schirmacher2021_disorder_classification}. Its precise definition and ways to probe it will be discussed in detail below.

Another quantifier of mechanical disorder has been related to the spectral broadening of low-frequency phononic bands in solids~\cite{phonon_widths}. Low-frequency phonons exist in solids due to a broken global continuous symmetry, independently of whether the solids are ordered (e.g., crystalline) or disordered. For finite-size solids, lowest-frequency phonons appear in well-separated, discrete bands. If the solid is ordered, the discrete phononic bands are degenerate, i.e.~groups of phonons with different wavevectors share the same frequency $\omega$. In the presence of mechanical disorder, this degeneracy is lifted and the discrete phononic bands acquire a finite spectral width $\Delta\omega$, as demonstrated in Fig.~\ref{fig:illustration_fig}. Consequently, $\Delta\omega/\omega$ can be used to construct a dimensionless disorder quantifier. The precise definition based on $\Delta\omega/\omega$ and ways to probe it will be discussed in detail below. The main questions we aim at addressing in this paper are whether the disorder quantifier defined based on $\Delta\mu/\langle\mu\rangle$ is fundamentally related to the one defined based on $\Delta\omega/\omega$, and if so, what the physical content and meaning of such a basic relation are.

To better understand these questions and to sharpen them, we present in Fig.~\ref{fig:illustration_fig}a the low-frequency vibrational density of states ${\cal D}(\omega)$ (VDoS) of finite-size computer glasses comprised of $N\!=\!\mbox{64,000}$ particles in three dimensions, obtained by quenching a deeply supercooled equilibrium liquid to zero temperature (details about the model and methods are provided below). It is observed that phononic bands localized at discrete frequencies exist at the lowest tail of the VDoS. Each phononic band is also characterized by a well defined width $\Delta\omega$, explicitly marked on the third band for illustration, making the ratio $\Delta\omega/\omega$ well-defined for each band. The lowest phononic bands are shown to be superimposed on top of an $\omega^4$ function (continuous green line), as highlighted by the zoom-in view presented in Fig.~\ref{fig:illustration_fig}b. The $\omega^4$ law corresponds to nonphononic excitations, which have been recently shown to be a universal feature of glasses formed by quenching a melt~\cite{modes_prl_2016,ikeda_pnas,modes_prl_2018,LB_modes_2019,modes_prl_2020}. As the nonphononic part of the VDoS universally follows the $\omega^4$ law, the prefactor $A_{\rm g}$ of this law (see figure) encapsulates nonuniversal properties of the glass, and in particular must be sensitive to the degree of mechanical disorder, which depends on the glass preparation procedure~\cite{cge_paper,LB_modes_2019,pinching_pnas}. 

At higher frequencies, the VDoS changes its character. In particular, when a crossover frequency $\omega_\dagger$ (which depends on the linear system size $L$) is surpassed, discrete phononic bands are no longer clearly distinguishable~\cite{phonon_widths}. Moreover, the $\omega^4$ law of nonphononic excitations is not clearly observed due to the overwhelming abundance of phonons, which eventually (at frequencies $\omega\!>\!\omega_\dagger$) follow Debye's density of states $\sim\!\omega^{\dbar-1}$ in $\dbar$ spatial dimensions (not shown). In these terms, the posed challenge is to understand how $\Delta\mu/\langle\mu\rangle$ manifests itself within the ``phonon sea'' for $\omega\!>\!\omega_\dagger$ and how it might be related to the spectral broadening of discrete phononic bands at $\omega\!<\!\omega_\dagger$. If this challenge is met, then one is able to unify two apparently distinct dimensionless quantifiers of mechanical disorder, which are accessible in both finite-size computer solids and in macroscopically large solids.  In the latter context, one should also establish experimental procedures to probe the disorder quantifier.

In this work, a basic relation between the two mechanical disorder quantifiers discussed above is derived and substantiated through computer simulations. The former is achieved using two alternative and complementary theoretical approaches that intimately link mechanical disorder to the low-frequency vibrational spectra of solids and to its effect on wave attenuation in such systems. The relation to low-frequency vibrational spectra, following Fluctuating Elasticity Theory~\cite{Schirmacher_2006,Schirmacher_prl_2007,schirmacher2011comments,schirmacher_ruocco_arXiv_2020,Schirmacher2021_disorder_classification} and the very recent support it received~\cite{scattering_letter_geert}, makes the emerging unified disorder quantifier experimentally accessible using techniques such as neutron and inelastic X-ray scattering. The unified disorder quantifier is also applied to packings near the unjamming transition and to disordered elastic spring-networks. 

Consequently, our work theoretically and computationally substantiates a unified quantifier of mechanical disorder in solids, thus allowing to quantify mechanical disorder across widely different physical systems using various probing techniques. At the same time, it offers insight into the low-frequency vibrational spectra of disordered solids in general, and in particular those of glassy systems.

\section{Observables and theory}
\label{sec:theory}

In this Section we introduce and relate two broadly applicable quantifiers of mechanical disorder, both illustrated in Fig.~\ref{fig:illustration_fig}. 
The first quantifier, $\chi$, is related to the spectral broadening of discrete phonon bands that emerge at the lowest frequencies of the vibrational spectrum of finite-size solids, as marked in Fig.~\ref{fig:illustration_fig}a. The second quantifier is known as the `disorder parameter' $\gamma$~\cite{Schirmacher_2006,Schirmacher_prl_2007,schirmacher2011comments,schirmacher_ruocco_arXiv_2020,Schirmacher2021_disorder_classification}, and is related to the relative width of the spatial or sample-to-sample~\cite{footnote} distribution of the shear modulus $\mu$, as shown for example in Fig.~\ref{fig:illustration_fig}d. 

\subsection{The mechanical-disorder quantifier $\chi$}

Phonons are collective, wave-like excitations that emerge in solids due to global translational symmetry~\cite{ashcroft1976solid}. In an ideal homogeneous and isotropic linear-elastic medium, phonons that share the same wavelength feature the exact same vibrational frequency, if they are also of the same polarization (shear or sound waves). This frequency degeneracy of phonons, expected in ideal continua, is lifted in disordered solids due to structural and mechanical disorder~\cite{phonon_widths}. Consequently, ideal phononic modes of a particular wavevector $\kv$ and frequency $\omega$ feature sizable projections on eigenvectors (normal modes) of the disordered system that are characterized by a frequency range $\Delta\omega$. We hereafter refer to $\Delta\omega$, which corresponds to the spectral width of the dynamic structure factor at a wavevector $\kv$, as the \emph{spectral width} pertaining to the wavevector $\kv$. In an isotropic solid, the spectral widths depend on the magnitude $k\!\equiv\!|\kv|$ (in addition to other observables, see below); in what follows, we restrict the discussion to isotropic media.

How does the spectral width $\Delta\omega(k)$ depend on $k$? At the smallest allowed phonon frequencies of a solid of linear size $L$, the spectral widths $\Delta\omega(k)$ are smaller than the frequency gaps between phonons of successive allowed wavevectors, i.e.~they remain \emph{discrete}, as is clearly shown in Fig.~\ref{fig:illustration_fig}a. Sorting the low-frequency, discrete phonon bands by increasing wavevectors $k$, we denote the phonon band index by $z$, and the (lifted) degeneracy of the $z$'th band by $n_z$. The latter is given by the number of \emph{different} solutions to the sum-of-integer-squares problem, namely the number of different combinations of integers $m_x,m_y,m_z$ such that
\begin{equation}
    m_x^2 + m_y^2 + m_z^2 = z\,.
\end{equation}
Notice that the wavevector $\kv$ is related to the integers $\mv\!\equiv\!(m_x,m_y,m_z)$ as $\kv\!=\!(2\pi/L)\mv$~\cite{ashcroft1976solid}, and $|\mv|\!=\!\sqrt{z}$. 

In~\cite{phonon_widths}, a perturbation theory was developed, culminating with a prediction for the scaling behavior of the spectral widths of discrete phonon bands; it reads~\cite{phonon_widths}
\begin{equation}\label{eq:phonon_band_widths}
    \Delta\omega\big(\omega,z,N\big) \propto \frac{\omega \sqrt{n_z}}{\sqrt{N}}\,,
\end{equation}
where $N$ is the system size, $\omega\!\simeq\!ck$ is the frequency of phonons from the $z$'th band ($c$ is the wave speed), and $n_z$ is their (lifted) degeneracy level. The prefactor of this proportionality relation defines a dimensionless mechanical disorder quantifier $\chi$, namely~\cite{phonon_widths}
\begin{equation}
\label{eq:chi_fs_definition}
  \chi \equiv \frac{\Delta\omega\big(\omega,z,N\big)\sqrt{N}}{\omega\sqrt{n_z}}\,.
\end{equation}
Figures~\ref{fig:illustration_fig}a,c present the low-frequency VDoS of computer glasses of $N\!=\!\mbox{64,000}$ quenched from different equilibrium parent temperatures $T_{\rm p}$ (as indicated by the legends); the different degrees of mechanical disorder featured by these glasses is manifested in the much-larger spectral widths $\Delta \omega$ of the high-$T_{\rm p}$ glasses. 

\subsection{The `phonon sea' crossover frequency $\omega_\dagger$}

Equation~(\ref{eq:phonon_band_widths}) for the spectral widths of discrete phonon bands holds up to a system-size dependent crossover frequency scale $\omega_\dagger(L)$, defined by the condition that the spectral width of discrete phonon bands becomes comparable to the gaps between successive phonon bands. Incorporating the definition of $\chi$ spelled out above and following~\cite{phonon_widths}, the crossover frequency $\omega_\dagger$ is predicted to satisfy~\cite{phonon_widths}
\begin{equation}
\label{eq:omega_dagger}
    \omega_\dagger \sim \big(\chi L\big)^{\frac{2}{\dbar+2}}\,,
\end{equation}
where $\dbar$ again denotes the dimension of space. $\omega_\dagger$ is marked by the vertical line in the example of Fig.~\ref{fig:illustration_fig}a, and its scaling with $L$ was validated using numerical simulations in~\cite{scattering_jcp}. At frequencies $\omega\!>\!\omega_\dagger$, discrete phonon bands are no longer distinguishable, and a `phonon sea' emerges instead, as seen in Figs.~\ref{fig:illustration_fig}a,c. Since $\omega_\dagger\!\to\!0$ in the thermodynamic limit, the phonon sea is expected to extend to zero frequency in that limit. In the phonon sea, Debye's VDoS~\cite{ashcroft1976solid} for phonon frequencies ${\cal D}(\omega)\sim\omega^{\dbar-1}$ is expected to hold.

\subsection{Spectral widths in the phonon sea}

How do the spectral widths $\Delta\omega(k)$ behave at frequencies $\omega\!>\!\omega_\dagger$, deep inside the phonon sea? Let us assume that Eq.~(\ref{eq:phonon_band_widths}) for the spectral width of discrete phonon bands with $\omega\!<\!\omega_\dagger$ also holds in the phonon sea; since the (lifted) degeneracy of phonons is no longer relevant (phonon bands are no longer discrete and well-separated in frequency), now $n_z$ represents the \emph{number} of modes that exist within the spectral widths $\Delta\omega$ (recall the physical meaning of the latter, as discussed above). $n_z$ can therefore be related to the vibrational density of states ${\cal D}(\omega)$ via~\cite{scattering_jcp}
\begin{equation}
    n_z \simeq N{\cal D}(\omega)\Delta\omega\,.
\end{equation}
Using this relation together with Eq.~(\ref{eq:phonon_band_widths}), we obtain
\begin{equation}\label{eq:rayleigh_klemens_law}
    \Delta\omega \simeq \chi^2\omega^2{\cal D}(\omega)\,,
\end{equation}
which, importantly, is independent of the system size $N$. Equation~(\ref{eq:rayleigh_klemens_law}) is closely related to the ``Rayleigh-Klemens law”~\cite{Klemens_1951}. In the thermodynamic limit, we expect phonons to dominate the low-frequency spectrum~\cite{phonon_widths}, namely ${\cal D}(\omega)\!\sim\!\omega_0^{-\dbar}\omega^{\dbar-1}$; we thus obtain the spectral widths inside the phonon sea in the form
\begin{equation}
\label{eq:scaling_theory_spectral_width}
    \Delta \omega \sim \chi^2\omega_0^{-\dbar}\omega^{\dbar+1}\,.
\end{equation}
In the context of the attenuation rate of plane waves in disordered media, the $\sim\!\omega^{\dbar+1}$ scaling is known as Rayleigh scattering~\cite{rayleigh_original}, and has been observed in numerical simulations in recent years~\cite{scattering_jcp,Ikeda_scattering_2018,wang2019sound,scattering_letter_geert,massimo_scattering_2021}. 

\subsection{The disorder parameter $\gamma$}

Fluctuating Elasticity Theory (FET)~\cite{Schirmacher_2006,Schirmacher_prl_2007,schirmacher2011comments,schirmacher_ruocco_arXiv_2020,Schirmacher2021_disorder_classification} is a theoretical framework that relates the spatial fluctuations of the local elastic moduli fields of a disordered solid to its vibrational and thermodynamic properties. Of particular interest here is the low-frequency wave attenuation rate $\Gamma$, which according to FET scales as $\gamma \omega_0^{-\dbar}\omega^{\dbar+1}$, where $\gamma$ is called the `disorder parameter' defined as
\begin{equation}\label{eq:gamma_definition}
    \gamma\equiv \left(\frac{\Delta\mu}{\langle\mu\rangle}\right)^2\left(\frac{\ell}{a_0}\right)^\dbar\,.
\end{equation}
Here $\ell$ represents the coarse-graining length on which the spatial fluctuations $\Delta\mu$ of the shear modulus are evaluated, and $a_0$ is an interparticle length as defined above. Assuming that the spatial distribution of the shear modulus field is correlated on a length scale $\xi_{\rm g}\!<\!\ell$, one expects the variance $(\Delta\mu)^2$ to scale as $\sim\!\ell^{-\dbar}$, and therefore $\gamma$ should become independent of the coarse-graining length $\ell$ for large enough $\ell$. Since we expect $\Gamma$ to be proportional to $\Delta \omega$, we conclude that the spectral width $\Delta\omega$ at frequency $\omega$ follows
\begin{equation}
\label{eq:fet_spectral_width}
    \Delta\omega \propto \gamma \omega_0^{-\dbar}\omega^{\dbar+1}\,.
\end{equation}

\subsection{A unified quantifier of mechanical disorder}

Combining now Eqs.~(\ref{eq:scaling_theory_spectral_width}) and (\ref{eq:fet_spectral_width}), we immediately conclude that 
\begin{equation}
\label{eq:main_result1}
    \chi^2 \sim \gamma\,,
\end{equation}
which is the main result of this work. 

The very same result can be obtained based on Eqs.~(\ref{eq:phonon_band_widths}) and (\ref{eq:fet_spectral_width}) alone; recall that the degeneracy $n_z(k)$ of wavevectors of magnitude $k$ (for a perfectly homogeneous solid) is proportional to $k^{\dbar-1}$~\cite{ashcroft1976solid}, and that $k\!\sim\!\sqrt{z}/L$. Therefore, $n_z\!\sim\! z^{\frac{\dbar - 2}{2}}\!\sim\!L^{\dbar - 2}\omega^{\dbar - 2}$~\cite{phonon_widths}, and hence for large $z$ phonon bands (at frequencies $\omega\!<\!\omega_\dagger$), we expect
\begin{equation}
	\Delta\omega \sim \frac{\chi \omega^{\dbar / 2}}{L}.
\end{equation}
Requiring that the spectral widths for $\omega\!<\!\omega_\dagger$ smoothly connect to the spectral widths at $\omega\!>\!\omega_\dagger$ as given by Eq.~(\ref{eq:omega_dagger}), we obtain
\begin{equation}
	\chi \frac{\omega_{\dagger}^{\dbar/2}}{L} \sim \gamma \omega_{\dagger}^{\dbar + 1} \quad\implies\quad
	\omega_{\dagger} \sim \qty(\frac{\gamma L}{\chi})^{-\frac{2}{\dbar+2}}\,.
\end{equation}
Comparing this result with the scaling relation for $\omega_{\dagger}$ in Eq.~\eqref{eq:omega_dagger}, we obtain
\begin{equation}
	(\chi L)^{-\frac{2}{\dbar+2}} \sim \qty(\frac{\gamma L}{\chi})^{-\frac{2}{ \dbar+2}} \quad\implies\quad \chi^2 \sim \gamma\,,
\end{equation}
in agreement with Eq.~(\ref{eq:main_result1}) above.

\section{Numerical support}
\label{sec:numerics}

Our goal in this Section is to test our main prediction in Eq.~(\ref{eq:main_result1}) for computer glasses generated over a broad range of conditions, which mimic a correspondingly large range of cooling rates through the glass transition. 

\subsection{Models and methods}

In this work, we employ a computer glass model of highly polydispersed soft spheres interacting via a $\propto\!r^{-10}$ pairwise potential, where $r$ is the distance between the centers of pairs of soft spheres. A full description of the model can be found in~\cite{boring_paper}. The model is inspired by the one put forward in~\cite{LB_swap_prx}, and as such it can be equilibrated down to very low temperatures using the swap-Monte-Carlo algorithm, where the latter is also explained in detail in~\cite{LB_swap_prx}. The crossover temperature of this system is found to be $\Tx\!\approx\!0.66$ in the model's simulation units (as described in~\cite{boring_paper}), according to the definition introduced in~\cite{crossover_temperature_jcp}. $\Tx$ coincides with the onset of the high-$T_{\rm p}$ plateau of $\gamma$ as shown in Fig.~\ref{fig:compare_chiz} below. The high-$T_{\rm p}$ shear modulus of this system is $\mu_\infty\!\approx\!9.2$ in the model's simulation units. 

We created ensembles of glassy samples of $N\!=\!\mbox{16,000}$ particles, parameterized by the equilibrium parent temperature $T_{\rm p}$, from which liquid configurations were instantaneously quenched using a conventional minimization algorithm. The sample-to-sample fluctuations of the shear modulus were evaluated using ensembles of 2,000 independent glassy samples for each $T_{\rm p}$. We also created a similar set of glass ensembles with $N\!=\!\mbox{2,000}$ particles, to demonstrate the strength of finite-size effects. For our spectral-widths calculations described below, it is necessary to employ somewhat large systems in order to cleanly estimate the widths of the lowest-frequency phonon bands in solids quenched from high parent temperatures. The system sizes we employed for these calculations, per each parent temperature, are described in the legend of Fig.~\ref{fig:phonon_width_N_Tp}a below. 


\subsection{Spectral widths $\Delta\omega$ of discrete phonon-bands}

Phonon band widths $\Delta\omega$ for each $T_{\rm p}$-ensemble were estimated as follows:
\begin{enumerate}
    \item We performed a partial diagonalization of the Hessian matrix $\bm{H}\!\equiv\!\frac{\partial^2U}{\partial\xv\partial\xv}$ for at least 50 independent configurations for each $T_{\rm p}$-ensemble to obtain the vibrational eigenfrequencies $\omega_\ell$ (all particle masses in our model are set to unity) and their associated eigenmodes $\bm{\Psi}^{(\ell)}$. 
    \item We filtered eigenfrequencies $\omega_\ell$ according to the participation ratio $e(\bm{\Psi}^{(\ell)})\!\equiv\!\big[N\sum_i({\bm\Psi}^{(\ell)}_i\!\cdot\!{\bm\Psi}^{(\ell)}_i)^2\big]^{-1}$ of their corresponding eigenmodes $\bm{\Psi}^{(\ell)}$, as also done in~\cite{ikeda_pnas,phonon_widths}. In our analyses, we only considered eigenmodes $\bm{\Psi}^{(\ell)}$ with $e(\bm{\Psi}^{(\ell)})\!>\!0.03$ in order to prevent low-frequency quasilocalized modes from affecting our estimations of phonon-band widths.
    \item We fitted a Gaussian to each peak pertaining to individual phonon bands, as done and explained in~\cite{phonon_widths}. The spectral widths $\Delta \omega$ were taken as the standard deviation (std) obtained from those Gaussian fits. 
\end{enumerate}

\begin{figure}[!ht]
\centering
\includegraphics[width = 0.5\textwidth]{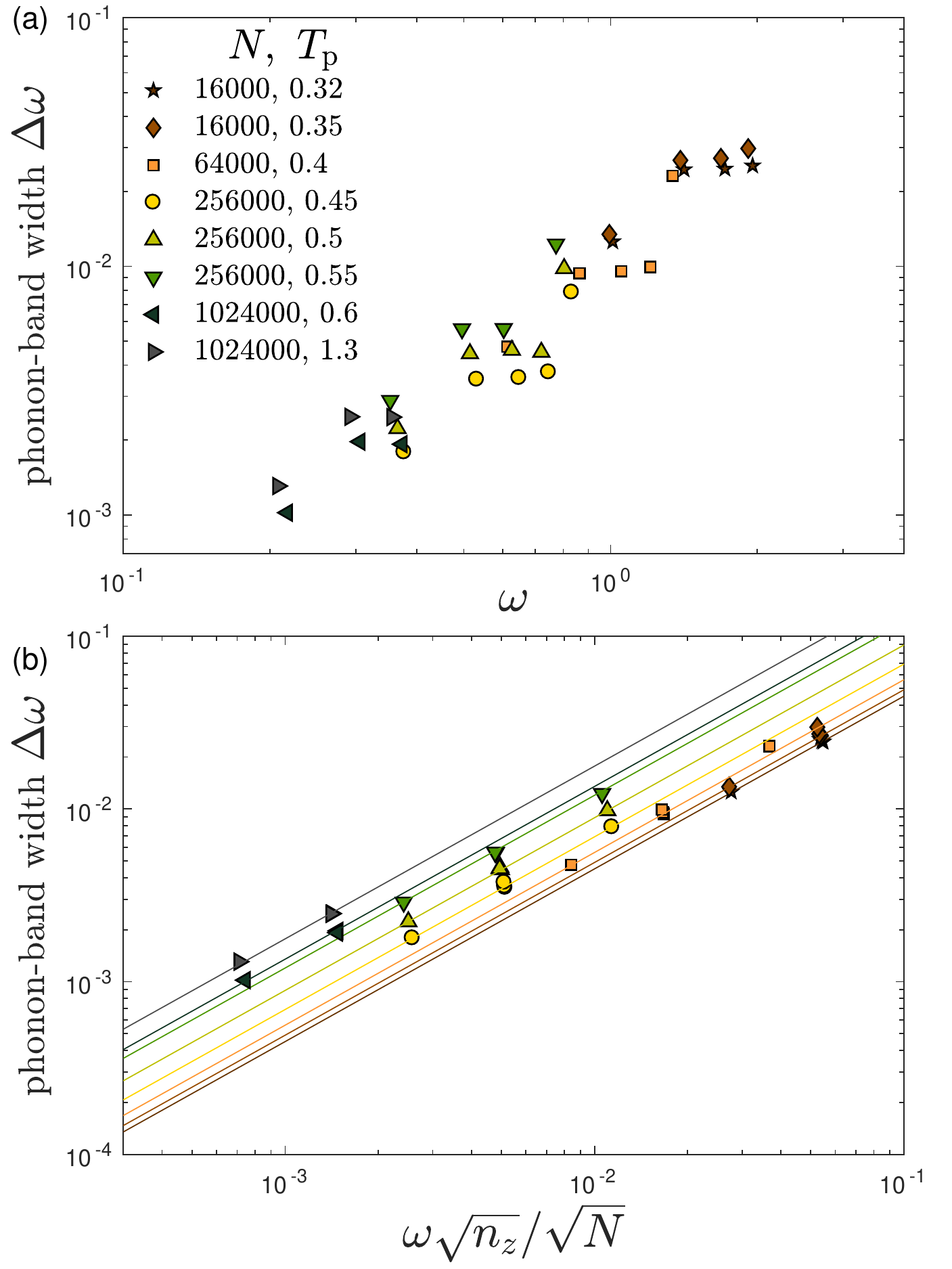}
\caption{\footnotesize (a) Phonon-band widths $\Delta\omega$ plotted against phonon bands frequencies $\omega$ (expressed in simulation units, see~\cite{boring_paper}). Data were measured for glasses of various sizes $N$ and parent temperatures $T_{\rm p}$ as indicated by the legend. (b)~Same phonon-band widths $\Delta\omega$ as reported in (a), this time plotted against the rescaled frequency $\omega\sqrt{n_z}/\sqrt{N}$. The continuous lines are fits to Eq.~(\ref{eq:chi_fs_definition}) from which we estimate $\chi(T_{\rm p})$.}
\label{fig:phonon_width_N_Tp}
\end{figure}

In Fig.~\ref{fig:phonon_width_N_Tp}, we present our measurements of our different $T_{\rm p}$ glasses' phonon-band widths $\Delta \omega$. In panel (a), we present the raw data obtained from the Gaussian fits as explained above. In panel (b), we plot the same data as shown in panel (a), this time against the rescaled frequency $\omega\sqrt{n_z}/\sqrt{N}$. For each $T_{\rm p}$, we fit the data following Eq.~(\ref{eq:phonon_band_widths}) --- represented by the continuous $\propto\!\omega$ lines in Fig.~\ref{fig:phonon_width_N_Tp}b --- to obtain an estimation of $\chi(T_{\rm p})$ in accordance with its definition in Eq.~(\ref{eq:chi_fs_definition}).

\subsection{Sample-to-sample $\mu$-fluctuations}
\label{sec:gamma_numerics}

\begin{figure}[!ht]
\centering
\includegraphics[width = 0.5\textwidth]{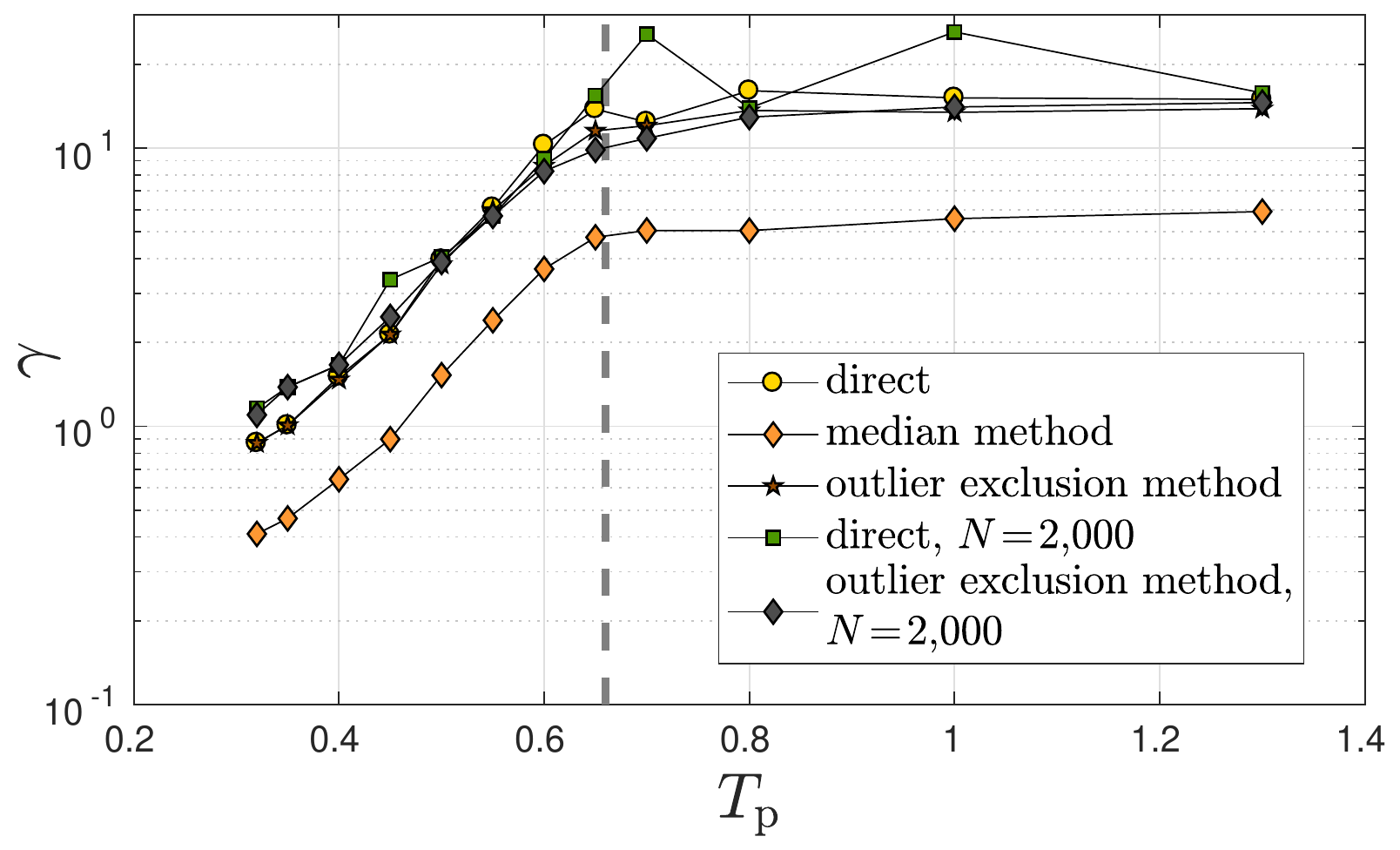}
\caption{\footnotesize Comparison between the direct calculation of $\gamma$ (via sample-to-sample statistics), and the two estimation methods as described in the text, for computer glasses of $N\!=\!\mbox{16,000}$ particles. The vertical dashed line marks the crossover temperature $\Tx$~\cite{crossover_temperature_jcp}, which coincides with the onset of the high-$T_{\rm p}$ plateau of $\gamma$. Also plotted are direct measurements and the outlier exclusion estimations of $\gamma$ for systems of $N\!=\!\mbox{2,000}$ particles (using the same ensemble-sizes as the $N\!=\!\mbox{16,000}$ systems), which are much noisier~\cite{karina_sticky1}. We find that, for $N\!=\!\mbox{16,000}$, the outlier exclusion method estimation follows closely the direct measurement at low $T_{\rm p}$, and is roughly 55\% higher than the median-based estimation throughout the studied $T_{\rm p}$-range. We therefore adopt the outlier-exclusion estimate of $\gamma$ in what follows.}
\label{fig:compare_chiz}
\end{figure}

Having at hand estimations for $\chi$, we now turn to estimating $\gamma$ via the sample-to-sample fluctuations $\Delta\mu$ of the shear modulus~$\mu$. Support for the equivalence between these procedures has been presented recently in~\cite{massimo_scattering_2021}. To this aim, we first stress that the sample-to-sample distribution $p(\mu;N)$ of glasses of size $N$ shows strong finite-size effects, as discussed at length in \cite{scattering_letter_geert,karina_sticky1}. In particular, in small glass samples quenched from high parent temperatures, some occurrences of anomalously large fluctuations of $\mu$ are often observed, rendering a clean estimation of the variance $(\Delta\mu)^2$ difficult. This finite-size effect --- which is also demonstrated in Fig.~\ref{fig:compare_chiz} below --- is likely related to the deviations from the $\omega^4$ law observed in small computer glasses instantaneously quenched from high parent temperatures~\cite{lerner2019finite}.

To overcome this potential difficulty, we adopt and compare between two approaches introduced in~\cite{karina_sticky1} and \cite{boring_paper} respectively:
\begin{enumerate}
    \item We follow the procedure described in~\cite{karina_sticky1} to remove outliers from each data set $\{\mu_i\}$ pertaining to each parent temperature $T_{\rm p}$, as follows:
    for each data point $\mu_i$, we calculate std($\mu$) for all \emph{other} data points $j\!\ne\!i$, namely under the \emph{exclusion} of $\mu_i$. We then identify the data point whose exclusion leads to the \emph{largest} variation of std($\mu$) amongst all other data points; if that (largest) variation with respect to the original (unfiltered) value of std($\mu$) exceeds 1\%, we permanently remove the identified data point from the total data set. This procedure is repeated until the variation of the standard deviation std($\mu$) under exclusion of any single data point is smaller than 1\%. This method is referred to below as the \emph{outlier exclusion} method. 
    \item In~\cite{boring_paper}, a measure of the width of the sample-to-sample distribution $p(\mu)$ was defined as the square-root of the \emph{median} (instead of the mean, as done for obtaining std($\mu$)) of the squared fluctuations $(\mu_i\!-\!\langle\mu\rangle)^2$ about the ensemble-mean $\langle\mu\rangle$. We refer to this method as the \emph{median} method. Notice that here we report the median of $(\mu_i\!-\!\langle\mu\rangle)^2$ rather than the square-root of the median as reported in~\cite{boring_paper}. 
\end{enumerate}

In Fig.~\ref{fig:compare_chiz}, we compare our estimations of $\gamma$ using the direct calculation and the two analysis schemes described above (outlier-exclusion and median methods). We find that the outlier-exclusion method results in slightly lower values of $\gamma$ for high $T_{\rm p}$ glasses. Importantly, we reiterate that our calculations are performed on ensembles of 2,000 glasses of $N\!=$ 16,000 particles, explaining why the difference between the direct calculation and the outlier elimination method are quite underwhelming. In Fig.~\ref{fig:compare_chiz} we also show the direct calculation of $\gamma$ for glasses of $N\!=\!\mbox{2,000}$ particles, which is substantially noisier, see~\cite{karina_sticky1} for a related discussion. We further note that the data obtained with the outlier exclusion method are roughly proportional to the estimations of $\gamma$ based on the median of fluctuations as described above, which is much less sensitive to outliers. For these reasons, we opt for the outlier-exclusion method to estimate $\gamma$ in what follows below. Finally, note that the crossover temperature $\Tx$~\cite{crossover_temperature_jcp}, marked by vertical dashed line in Fig.~\ref{fig:compare_chiz}, appears to coincide with the onset of the high-$T_{\rm p}$ plateau of $\gamma$. The physical significance and relevance of this interesting observation will be discussed elsewhere. 

\begin{figure}[!ht]
\centering
\vspace{0.5cm}
\includegraphics[width = 0.5\textwidth]{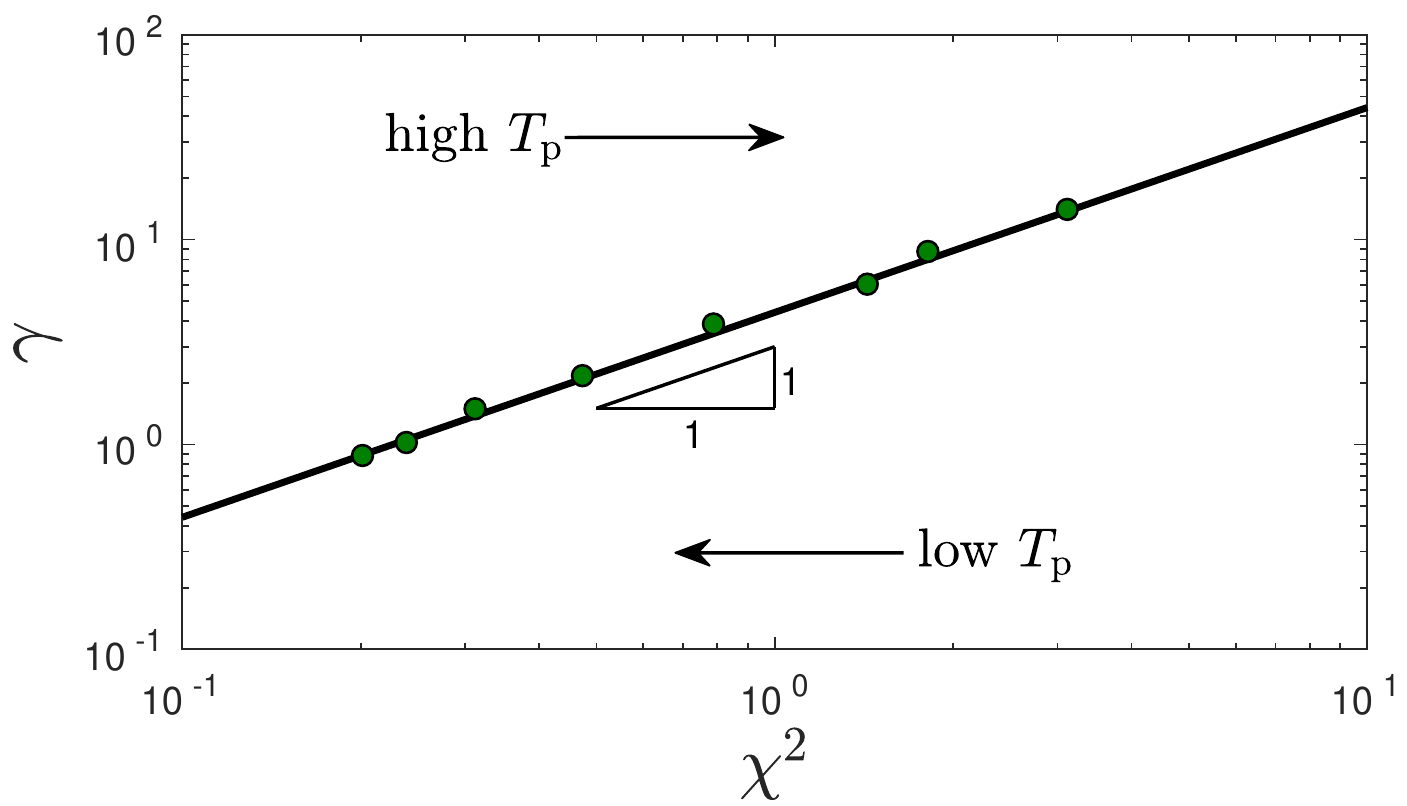}
\caption{\footnotesize Parametric plot of $\gamma(T_{\rm p})$ vs.~$\chi^2(T_{\rm p})$ obtained as explained in the text. The main result of this work is that $\gamma\!\sim\!\chi^2$, as predicted by our scaling theory in Eq.~(\ref{eq:main_result1}).}
\label{fig:chi_td_vs_chi_fs}
\end{figure}

With estimations of $\gamma(T_{\rm p})$ and $\chi(T_{\rm p})$ at hand, we parametrically plot the two quantifiers against each other in Fig.~\ref{fig:chi_td_vs_chi_fs} to find that $\gamma\!\sim\!\chi^2$, as predicted by our scaling theory in Sect.~\ref{sec:theory}.

\section{Mechanical disorder near the unjamming point}

Up to now we demonstrated the validity of our main prediction in Eq.~(\ref{eq:main_result1}) for computer glasses quenched from a melt, cf.~Fig.~\ref{fig:chi_td_vs_chi_fs}. Our prediction, however, is expected to be generally valid for a broader class of disordered solids. This is demonstrated in this Section. 

The unjamming transition is a mechanical instability observed upon decompressing athermal packings of frictionless soft spheres~\cite{ohern2003,liu_review,van_hecke_review,liu2011jamming}. Growing lengthscales and correlation volumes are known to emerge in these systems as their confining pressure is reduced towards zero~\cite{Silbert_prl_2005,breakdown,everyone,brian_jamming_lengths_prl_2017,atsushi_core_size_pre,liu_jamming_lengths_prl_2018}. The key microscopic parameter controlling the mechanical behavior of low-pressure soft-sphere packings is the coordination $Z$, which represents the number of interactions per particle. In particular, scaling laws of elastic moduli~\cite{matthieu_thesis,Ellenbroek_2009} and lengthscales~\cite{Silbert_prl_2005,breakdown,brian_jamming_lengths_prl_2017,atsushi_core_size_pre,liu_jamming_lengths_prl_2018} with the difference $\delta Z\!\equiv\!Z\!-\!Z_c$ are known to emerge, where the critical coordination $Z_c\!=\!2\dbar$ --- known as the Maxwell threshold~\cite{maxwell_1864} --- is reached in the limit of vanishing confining pressure. 

How do the interrelated mechanical disorder quantifiers discussed here behave near the unjamming point? In~\cite{mw_emt_2010,eric_boson_peak_emt} the spectral widths of acoustic excitations are obtained using Effective Medium calculations on disordered Hookean spring networks of coordination $Z$; the key result relevant to the present discussion is the scaling $\Delta\omega\!\sim\!\omega^4/\delta Z^{5/2}$ (in three dimensions), implying together with Eq.~(\ref{eq:fet_spectral_width}) and $\omega_0\!\sim\!\sqrt{\mu}\!\sim\!\sqrt{\delta Z}$~\cite{Ellenbroek_2009} that $\gamma\!\sim\!1/\delta Z$. This result is consistent with numerical data from simulations of soft-sphere packings put forward in~\cite{everyone}, where it was shown that the sample-to-sample shear modulus distribution collapses for different pressures and system sizes if it is considered for systems with constant $N\delta Z$, implying again that $\gamma\!\sim\!1/\delta Z$ near the unjamming point.

\begin{figure}[!ht]
\centering
\includegraphics[width = 0.5\textwidth]{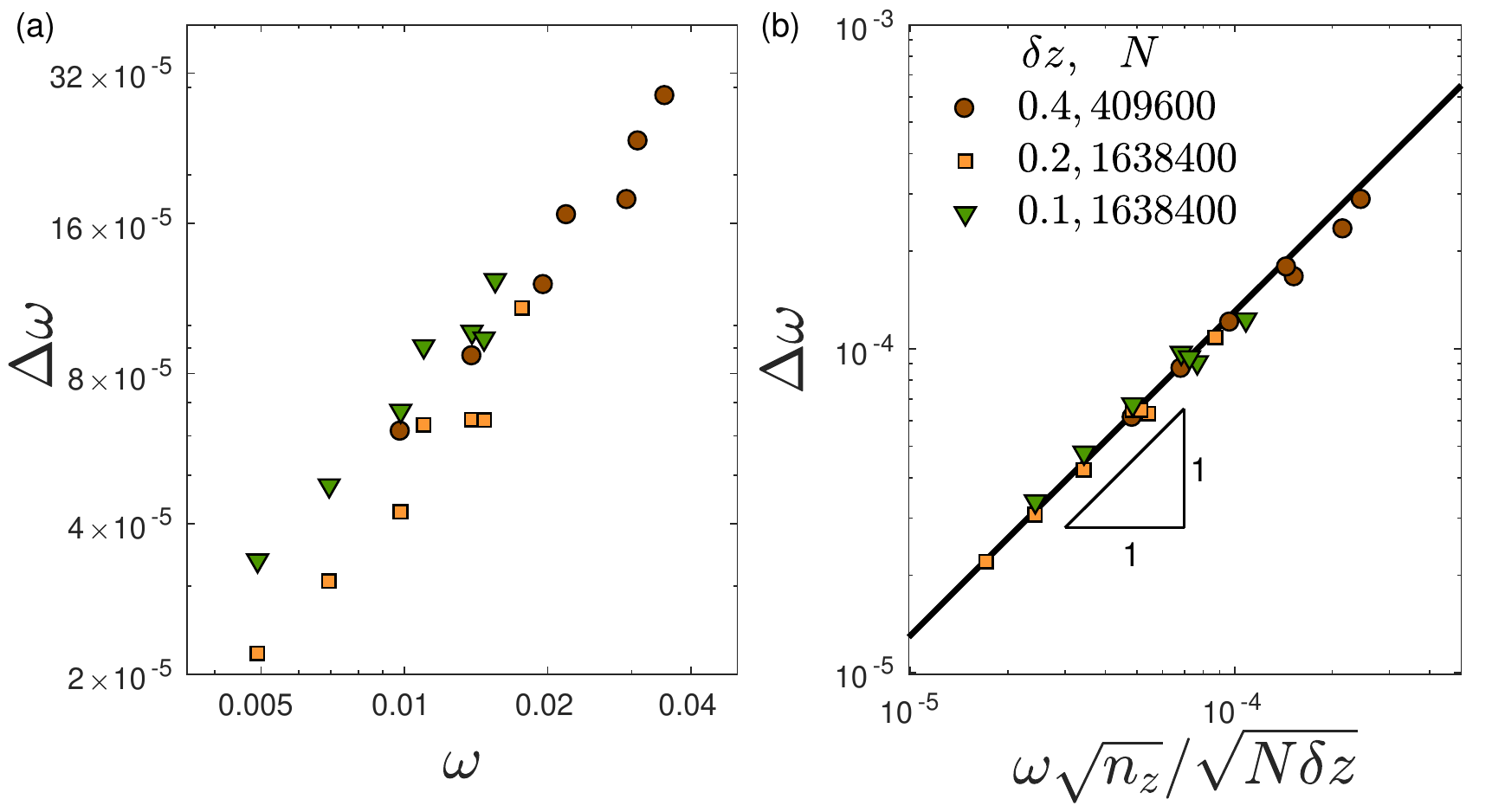}
\caption{\footnotesize Phonon-band widths $\Delta\omega$ measured in disordered spring networks in two-dimensions, see text and Appendix~\ref{sec:appendix} for details. $\Delta\omega$ is plotted against the frequency~$\omega$ in panel (a), and against the rescaled frequency $\omega\sqrt{n_z}/\sqrt{\delta ZN}\!\sim\!\chi\omega\sqrt{n_z}/\sqrt{N}$ in panel (b). The data collapse implies that $\chi\!\sim\!1/\sqrt{\delta Z}\!\sim\!\sqrt{\gamma}$, strengthening the generality of the scaling relation between the mechanical disorder quantifiers $\gamma$ and $\chi$.}
\label{fig:chi_in_spring_networks}
\end{figure}

If the relation $\gamma\!\sim\!\chi^2$ is general, it should also hold in systems near the unjamming transition as well. Here, we test this scaling relation by measuring the broadening of discrete, low-frequency phonon-bands of disordered spring networks in two-dimensions. Our networks' geometry was first derived from the contact-network of disordered soft-disc packings, and their coordination $Z$ was reduced towards $Z_c$ using a edge-dilution scheme described in Appendix~\ref{sec:appendix}, which minimizes fluctuations of the angles formed by edges around nodes. The results are presented in Fig.~\ref{fig:chi_in_spring_networks}; we find that $\chi\!\sim\!1/\sqrt{\delta Z}$, in agreement with our main theoretical prediction. 

\begin{figure*}[!ht]
\centering
\includegraphics[width = 1.0\textwidth]{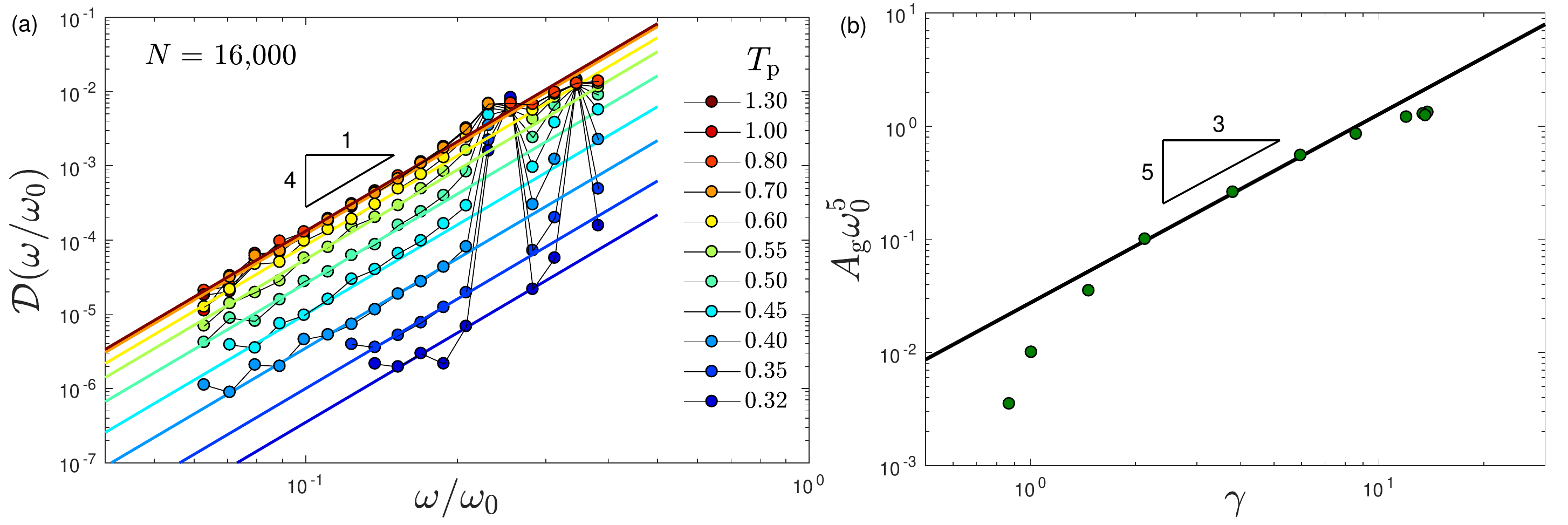}
\caption{\footnotesize (a) VDoS of our computer glasses of various $T_{\rm p}$ as indicated by the legend. The dimensionless prefactors $A_{\rm g}\omega_0^5$ are estimated by fitting the low frequency tails to the universal quartic law $\propto\!\omega^4$ of the nonphononic VDoS. The extracted dimensionless prefactors $A_{\rm g}\omega_0^5$ are then plotted in panel (b) vs.~the disorder parameter $\gamma$ obtained from the sample-to-sample fluctuations of the shear modulus as explained in the text. In~\cite{david_fracture_paper_arXiv} it is argued that $A_{\rm g}\omega_0^5\!\sim\!\gamma^{5/3}$, as represented by the continuous line.}
\label{fig:Ag_vs_gamma}
\end{figure*}

\section{Concluding remarks and prospects}
\label{sec:conclusions}

In this work, we have discussed two broadly-applicable and dimensionless quantifiers of mechanical disorder. The first quantifier, $\chi$, is related to the spectral broadening of discrete phonon bands seen in the low-frequency spectra of finite-size computer glasses, as shown in Fig.~\ref{fig:illustration_fig}a,c. The second quantifier, $\gamma$, is known as the `disorder parameter' in Fluctuating Elasticity Theory~\cite{Schirmacher_2006,Schirmacher_prl_2007,schirmacher2011comments,schirmacher_ruocco_arXiv_2020,Schirmacher2021_disorder_classification}, and plays a key role in determining the spectral widths of acoustic excitations in the thermodynamic limit. Our main result --- $\gamma\!\sim\!\chi^2$ --- was validated (in Fig.~\ref{fig:chi_td_vs_chi_fs}) against extensive computer simulations of glasses quenched from a broad range of parent temperatures $T_{\rm p}$. It was also validated (in Fig.~\ref{fig:chi_in_spring_networks}) against computer simulations of disordered spring networks approaching the unjamming transition. 

In order to assess the value of $\gamma$ --- which characterizes the relative width of the distribution of coarse-grained elastic moduli \emph{fields} in disordered media (see Eq.~(\ref{eq:gamma_definition})) --- we employed sample-to-sample statistics instead of spatial coarse-graining procedures. The equivalence of these two approaches to evaluating $\gamma$ was recently argued for in~\cite{massimo_scattering_2021} based on detailed analyses of computer glasses. A deeper understanding of this equivalence and its further reinforcement is left for future investigations. 

In Fig.~\ref{fig:illustration_fig}a-c, we presented data that indicate that the prefactor $A_{\rm g}$ of the universal $\propto\!\omega^4$ nonphononic VDoS correlates with the mechanical disorder quantifiers $\gamma$ and $\chi$ discussed in this work. It is natural to expect that systems rich with soft nonphononic modes --- as indicated by a large (dimensionless) prefactor $A_{\rm g}\omega_0^5$ --- would also have relatively large spatial fluctuations of their coarse-grained shear modulus fields. Indeed, in~\cite{david_fracture_paper_arXiv} it was argued that $A_{\rm g}\omega_0^5\!\sim\!\gamma^{5/3}$ in three dimensional glasses~\cite{footnote2}, based on scaling arguments. 

These predictions are tested against our computer glasses data in Fig.~\ref{fig:Ag_vs_gamma}. In panel (a), we show the low-frequency VDoS of our computer glasses of different parent temperatures $T_{\rm p}$; we obtain estimations of the dimensionless prefactors $A_{\rm g}\omega_0^5$ by fitting the low-frequency tails to the universal $\sim\!\omega^4$ law. In Fig.~\ref{fig:Ag_vs_gamma}b, we plot the extracted $A_{\rm g}\omega_0^5$ vs.~the disorder parameter $\gamma$ obtained as explained in Sect.~\ref{sec:gamma_numerics}. We find that the proposed scaling $A_{\rm g}\omega_0^5\!\sim\!\gamma^{5/3}$ holds over a limited, intermediate range of $\gamma$ values. As pointed out in~\cite{karina_sticky1,minimal_disorder_arXiv}, $A_{\rm g}\omega_0^5$ dips downwards at the lowest values of $\gamma$. In~\cite{minimal_disorder_arXiv}, it was demonstrated that $\gamma$ varies monotonically with a glassy correlation length $\xi_{\rm g}$, which is expected to be bounded from below by an interparticle distance, suggesting a lower bound on $\gamma$ as well. Relations between the glassy length $\xi_{\rm g}$, the disorder parameter $\gamma$ and the prefactor $A_{\rm g}$ were suggested, discussed and tested further in~\cite{atsushi_core_size_pre,karina_sticky1,minimal_disorder_arXiv,massimo_scattering_2021}. A complete understanding of the relation between $A_{\rm g}$ and the disorder quantifiers discussed here is left for future work.

We conclude the discussion with commenting on the experimental accessibility of the mechanical disorder quantifiers discussed in this paper. Spectral widths of acoustic excitations $\Delta\omega(k)$ are related to wave attenuation rates $\Gamma(k)$ as $\Gamma\!\sim\Delta\omega$, as demonstrated for frequencies $\omega\!<\!\omega_\dagger$ using computer simulations in \cite{phonon_widths}, and for frequencies $\omega\!>\!\omega_\dagger$ in \cite{massimo_scattering_2021}. While longitudinal wave attenuation rates are accessible experimentally~\cite{ruta2012acoustic,baldi2014anharmonic} via high-resolution inelastic x-ray scattering, methods for measuring transverse (shear) waves' attenuation rates well in the Rayleigh scaling regime ($k\!\lesssim\!1$nm$^{-1}$) are unfortunately not yet available. In \cite{Wagner2011}, a measure of local elastic moduli of a metallic glass (amorphous PdCuSi) was obtained using atomic force acoustic microscopy. There, it was reported that $\Delta\mu/\mu\!\approx\!30\%$, where the fluctuations were estimated over lengths of order 10nm. An important goal of future experiments is to assess and compare the mechanical disorder of laboratory glasses to our measurements of the mechanical disorder of computer glasses on equal footing, in terms of the quantifiers discussed in this work. An impressive effort in this direction was presented very recently in \cite{Schirmacher2021_disorder_classification}.

\acknowledgements
We thank Jacques Zylberg for developing and sharing the edge-dilution algorithm that inspired the one used in this work. E.B.~acknowledges support from the Ben May Center for Chemical Theory and Computation and the Harold Perlman Family. E.L.~acknowledges support from the NWO (Vidi grant no.~680-47-554/3259).
\vspace{-0.2cm}

\appendix

\section{Disordered spring networks}
\label{sec:appendix}

\begin{figure}[!ht]
	\centering
	\includegraphics[width = 0.51\textwidth]{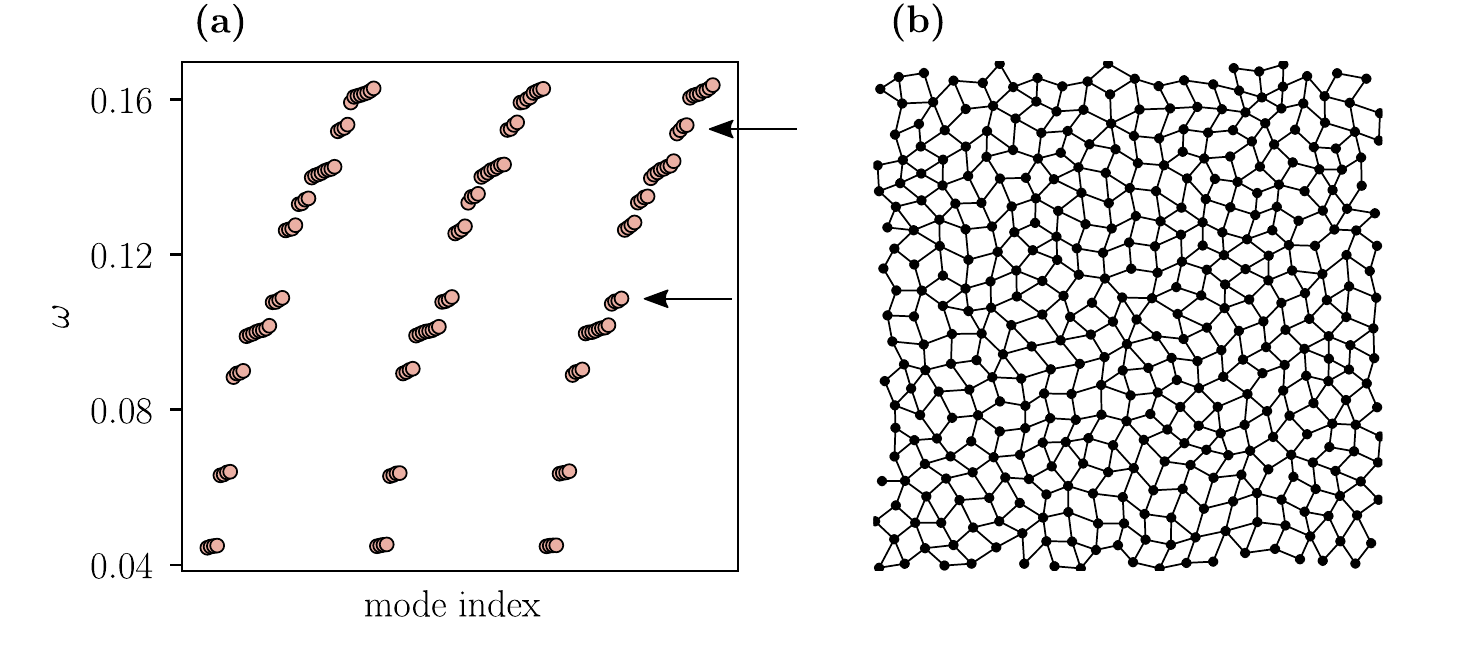}
	\caption{\footnotesize (a) Lowest-frequency phonon bands of three realizations of spring networks of $N\!=\!\mbox{1,638,400}$ nodes with coordination $Z\!=\!4.1$ generated using the algorithm described in the text. Sound waves are marked by arrows. (b) An example of a spring network of $N\!=\!400$ nodes and coordination $Z\!=\!4.1$ generated with the algorithm described here.}
	\label{fig:network_phonon_bands}
\end{figure}

We created two-dimensional disordered networks of Hookean springs by adopting the contact networks of the soft-disc glasses described and studied in~\cite{cge_paper}. The particle-centers of the original glass are set to be the networks' nodes, and an edge is placed between each pair of interacting particles in the original glass. The coordination of these initial networks is $Z\!\approx\!6.5$. We employed the edge-dilution algorithm described below to remove edges until the target coordinations $Z\!=\!4.4,4.2,4.1$ were reached. Each edge of the remaining network is then replaced by a \emph{relaxed} Hookean spring (of unit stiffness), i.e.~the spring's rest-length is set to the original distance between the pair of nodes it is connected to. An example of a network produced by our algorithm is shown in Fig.~\ref{fig:network_phonon_bands}b.

The low-frequency spectrum of the disordered networks produced by our scheme --- provided the system is large enough --- consists solely of phonons, as demonstrated in Fig.~\ref{fig:network_phonon_bands}a. This is a non-trivial feature of our scheme. We tested other edge-dilution schemes aimed at minimizing local coordination fluctuations. The low frequency spectrum of spring networks with small $\delta Z$ produced by these other schemes usually features localized, nonphononic soft vibrations, in addition to phonons. Since the low-frequency spectrum of the spring networks produced by our algorithm consists solely of phonons, we could directly measure the spectral width of phonon bands by calculating the standard deviation of the vibrational frequencies of each band.

\subsection{Bond-dilution algorithm}

\begin{figure}[!ht]
 	\centering
 	\includegraphics[width = 0.5\textwidth]{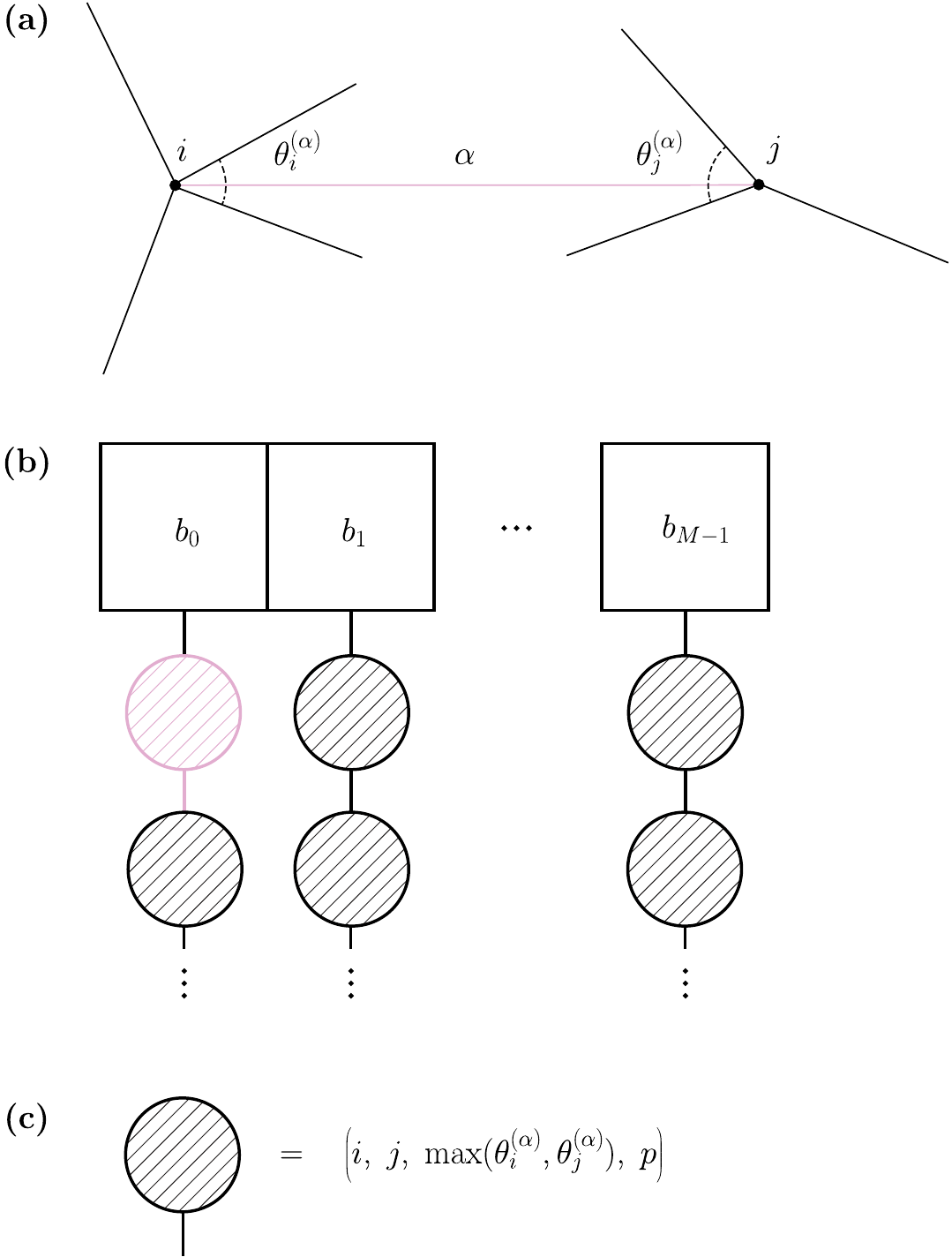}
	\caption{(a) When an edge $\alpha$ is removed, two remaining angles $\theta_i^{(\alpha)}$ and $\theta_j^{(\alpha)}$ are formed at nodes $i$ and $j$. In each iteration of the algorithm we aim to remove the edge whose removal forms the \emph{smallest} remaining angles. (b) Illustration of the key data structure of the algorithm; upon initialization, all edges of the initial network are stored in $b$ --- an array of linked lists --- so that $b_i$ contains a linked list of edges with $\frac{2\pi i}{M}\! <\! \max(\theta_i^{(\alpha)}, \theta_j^{(\alpha)})\!<\! \frac{2\pi (i + 1)}{M}$. The edge targeted for removal is shown in pink. (c) An element of a linked list contains an edge $\alpha$ specified by the two nodes $i$ and $j$, the largest bond angle that would open up in case of its removal, and a pointer $p$ to the next element of the linked list.}
 	\label{fig:angle_protocol}
 \end{figure}

We introduce an algorithm~\cite{zylberg_ack} that iteratively removes edges from a two-dimensional network, based on the network's geometry. To explain how the algorithm works, we refer readers to the illustration in Fig.~\ref{fig:angle_protocol}a; removing the edge labeled $\alpha$ between two nodes $i$ and $j$ creates two \emph{remaining} angles $\theta_i^{(\alpha)}$ and $\theta_j^{(\alpha)}$. In each edge-removal iteration of our algorithm, the next edge to be removed is the one whose larger (out of two) associated remaining angle (as defined above) is \emph{minimal}, across all edges.


The algorithm consists of a preprocessing step and a removal step. The preprocessing step creates an array of $M$ linked-lists, denoted by $b$. The links of each linked-list represent edges; each link representing an edge stores the larger of the two remaining angles that would be formed upon removal of that edge, namely $\theta^{(\alpha)}\!\equiv\! \max(\theta_i^{(\alpha)}, \theta_j^{(\alpha)})$. Each linked-list $b_\ell$ holds the edges $\alpha$ whose $\theta^{(\alpha)}$ are equal up to $2\pi/M$, namely those that satisfy $\frac{2\pi \ell}{M}\! <\! \theta^{(\alpha)}\! <\! \frac{2\pi (\ell + 1)}{M}$. This data structure is explained in Fig.~\ref{fig:angle_protocol}(b).

To find the edge that opens up the smallest bond-angle (up to accuracy $2\pi / M$), we simply find the first element of $b$ that contains a non-empty linked list, and remove the edge that is stored in the head of that linked list (shown in pink in Fig.~\ref{fig:angle_protocol}(b)). The second element of the linked list becomes its new head. To remove subsequent edges, we simply repeat this procedure, emptying the linked lists in the array $b$ from left to right.

Importantly, when a link is considered for removal, it is necessary to check if the originally stored $\theta^{(\alpha)}$ is still accurate, since previous edge-removals might have increased it. If $\theta^{(\alpha)}$ is unchanged, we remove the edge. If $\theta^{(\alpha)}$ has changed due to other edge-removals, we remove its link and reinsert it at the head of the linked list in $b$ corresponding to the updated remaining angle $\theta^{(\alpha)}$. The reason we can traverse $b$ from left to right is that $\theta^{(\alpha)}$ can only \emph{increase} when we remove other edges $\ne\!\alpha$. To create the networks used in this work, we chose $M\!=\!10^4$.

\begin{thebibliography}{53}
\bibitem [{zyl()}]{zylberg_ack}%
  \BibitemOpen
  \bibinfo {note} {Our algorithm is an adaptation of an algorithm developed by
  J.~Zylberg.}\BibitemShut {Stop}%
\end{thebibliography}

%

\end{document}